# Epsilon-near-zero plasmonic waveguides to enhance nonlinear coherent light-matter interactions

Ying Li, Christos Argyropoulos*

Department of Electrical and Computer Engineering, University of Nebraska-Lincoln, Lincoln, NE USA 68588

## ABSTRACT

We demonstrate a way to coherently control light at the nanoscale and achieve coherent perfect absorption (CPA) by using epsilon-near-zero (ENZ) plasmonic waveguides. The presented waveguides support an effective ENZ response at their cut-off frequency, combined with strong and homogeneous field enhancement along their nanochannels. The CPA conditions are perfectly satisfied at the ENZ frequency, surprisingly by a subwavelength plasmonic structure, resulting in strong CPA under the illumination of two counter-propagating plane waves with appropriate amplitudes and phases. In addition[*], we investigate the nonlinear response of the proposed ENZ plasmonic configuration as we increase the input intensity of the incident waves. We demonstrate that the CPA phenomenon can become both intensity- and phase-dependent in this case leading to new tunable all-optical switching and absorption devices.

## 1. INTRODUCTION

Perfect absorption was usually obtained under single beam illumination due to the critical coupling phenomenon [1,2]. Recently, a new mechanism to achieve perfect absorption, named coherent perfect absorption (CPA), was first theoretically introduced in 2010 [3] with a subsequent experimental verification in 2011 [4]. As the time-reversed counterpart of lasing, the CPA effect realized the interferometric control of absorption under the illumination of two counter-propagating coherent dual beams. Hence, it provided a new opportunity to modulate the absorption and control light in an efficient and coherent way [5]. This dynamic effect has been investigated in various photonic, plasmonic [6,7], and graphene structures [8,9]. In addition, CPA was also demonstrated in parity-time (PT) symmetric systems [10,11]. Spatially symmetric and balanced inclusions of loss and gain in photonic systems have led to PT-symmetric non-Hermitian systems that simultaneously behave as a coherent perfect absorber and a laser oscillator depending on the amplitude and phase of the two input beams (also known as CPA-laser [12]). However, currently, the CPA effect investigations are mostly focused on linear absorption effects. Only a few theoretical studies have begun to extend the CPA concept to nonlinear absorptive media [13–16].

Moreover, recently, realistic metamaterials exhibiting effective epsilon-near-zero (ENZ) permittivity response have raised increased attention, especially due to their peculiar transmission properties that provide, in principle, infinite phase velocity combined with an anomalous impedance-matching tunneling effect [17–19]. This leads to a quasi-static electromagnetic response that can be realized using narrow plasmonic waveguides operating at their cut-off wavelength. Uniform phase distribution and large field enhancement can be achieved inside the nanochannels of these narrow waveguides. These interesting properties have been used to enhance fluorescence, squeeze and tunnel light from bended waveguides, and boost nonlinear optical effects [20–23]. Interestingly, very recently, some works have been proposed that connect different ENZ metamaterial designs with the CPA effect [24–26]. However, none of these studies extends to the investigation of the nonlinear ENZ CPA response that promises to lead to a new degree of coherent nonlinear light-matter interactions.

In this work, we propose a plasmonic ENZ waveguide system to realize both linear and nonlinear CPA in the nanoscale. We first investigate the CPA operation in the linear plasmonic configuration case. The presented plasmonic devices exhibit an effective ENZ response at their cut-off frequency, as it was mentioned before, and Fabry-Pérot (FP) resonances at higher frequencies [21,22]. It will be demonstrated that perfect CPA can be achieved at the ENZ resonance under the illumination of two counter-propagating plane waves with appropriate amplitudes and phases. Then, we provide a novel way to turn on and off the CPA phenomenon by studying the Kerr third-order nonlinear effects arising due to the increased input intensities illuminating the proposed ENZ plasmonic waveguides. Finally, the influence of thermal effects is also scrutinized in order to provide further theoretical insights regarding the practicality of the proposed tunable CPA process. Our findings may have implications in nonlinear gap solitons [14], optical switches [15], ultrasensitive sensors, and tunable unidirectional coherent perfect absorbers [27].

---

[*]christos.argyropoulos@unl.edu; phone 1 402 472-3710; fax 1 402 472-4732; https://argyropoulos.unl.edu

## 2. LINEAR CPA WITH ENZ PLASMONIC WAVEGUIDES

We first consider the linear operation of the proposed plasmonic waveguides schematically shown in Fig. 1. The proposed plasmonic configuration is composed of periodic narrow rectangular slits carved in a silver (Ag) screen with complex relative permittivity dispersion following experimental values [28]. The slits are loaded with a linear dielectric material, named SF11 glass with relative permittivity of $\varepsilon_1 = 3.1$. The entire array of plasmonic waveguides is embedded in a dielectric host material which can be used as a substrate and superstrate to assist its practical implementation. In the currently investigated configuration, the substrate and superstrate are loaded with FK51 glass with permittivity $\varepsilon_1 = 2.2$. A cross-sectional view of one unit cell of the periodic waveguides is shown in the inset of Fig. 1, composed of a narrow rectangular slits of width $w$, height $t<<w$, nanochannels' length $l_1$, and periods $a$, $b>>t$. The presented free-standing plasmonic waveguide geometry to realize effective ENZ response was first introduced in [20]. The ENZ resonance of this structure can be tuned through properly modifying the lateral width $w$ of the nanochannels. The waveguide width $w$ is appropriately designed so the proposed structure will operate at the cutoff frequency of its dominant quasi-transverse electric (quasi-TE$_{10}$) mode, *i.e.*, at the ENZ frequency for which its guided wave number $\beta$ has a near-zero real part. Hence, the corresponding effective impedance of each waveguide becomes very large close to the cutoff frequency, since it is inversely proportional to the real part $Re[\beta]$ for TE modes. Such large modal impedance can be used to compensate the geometrical mismatch at the transition between each narrow nanowaveguide slit and surrounding dielectric materials. These properties result in an anomalous impedance matching phenomenon that is able to produce large coupling of the incident light inside each nanowaveguide, a combination of properties extremely well suited to boost the coherent light-matter interactions in the nanochannels. Such impedance matching phenomenon depends only on the aperture-to-period ratio of the array and is, therefore, independent of the channels' length or shape. Here, the dimensions of the rectangular channels are designed to have a waveguide length $l_1 = 700$ nm, substrate and superstrate layer thickness $l_2 = 350$ nm, period $a = b = 400$ nm, width $w = 200$nm, and height $t = 40$ nm ($t<<w$).

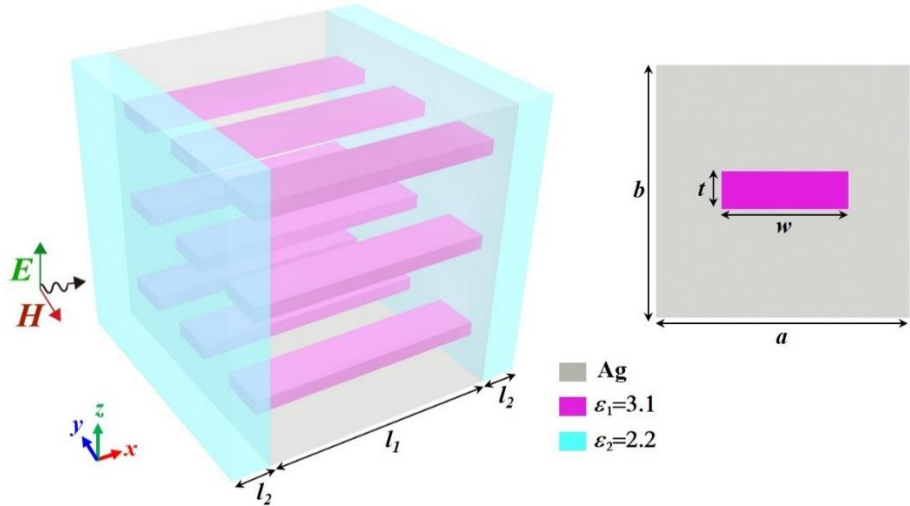

Figure 1. Geometry of the silver (Ag) plasmonic waveguides loaded with glass slits ($\varepsilon_1$=3.1). The array of waveguide channels is embedded inside a dielectric substrate and superstrate with $\varepsilon_2$=2.2. A cross-sectional view of one unit cell of the periodic silver nanowaveguides is presented in the right inset.

To prove the aforementioned interesting ENZ response, we compute the transmission and reflection coefficients of the proposed array of plasmonic waveguides and the results are shown in Fig. 2. In this case, the perforated metallic screen is illuminated by a normal incident *z*-polarized plane wave, as schematically illustrated in Fig. 1. There are two transmission peaks corresponding to two different supported plasmonic modes inside the nanochannels. Specifically, the dominant quasi-TE$_{10}$ mode in the metallic waveguides is brought to be at cutoff at $\lambda = 1202$ nm, where the effective electromagnetic properties of the plasmonic channels become equivalent to the propagation of an ENZ metamaterial. The computed field distribution in channel's *xy*-plane at the ENZ resonance demonstrates a uniformly enhanced electric field distribution due to the aforementioned ENZ response (inset of Fig. 2, upper right), independent of the total channel length, which can drastically enhance nonlinear effects by extending the effective interaction volume, in principle, indefinitely along the nanochannels. The plasmonic losses of the nanochannels are the only limiting factor in this interesting response. The electric field at a higher order resonance wavelength $\lambda = 1153$ nm exhibits usual FP resonant behavior and is also plotted in the inset of Fig. 2 (lower right). This resonance mode has the typical characteristics of a standing wave distribution, which is significantly different compared to the ENZ response. The red-dashed line shows the calculated phase difference $\Delta\phi$ between the transmission and reflection coefficients. Note that an abrupt phase change from almost $\pi$ to $-\pi$ is obtained at the ENZ resonance, a typical behavior of a resonating effect. We will explain later that either $\Delta\phi = 0$ or $\pi$ is one of the required conditions to generate CPA when the structure under study is illuminated by two counter-propagating plane waves.

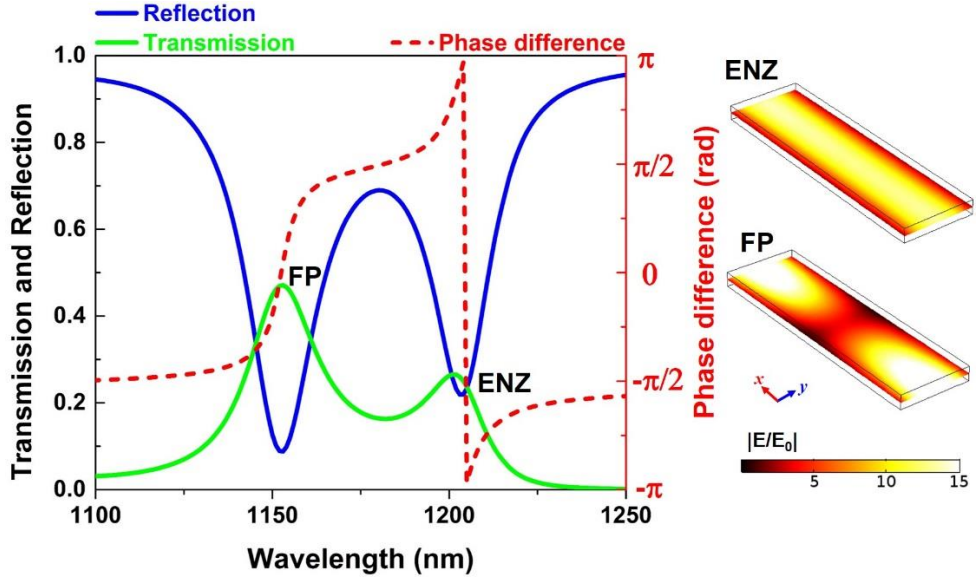

Figure 2. Computed transmission and reflection coefficients and phase difference of the ENZ plasmonic waveguide device as a function of the incident wavelength. Right inset: electric field enhancement distribution in the channel's *xy*-plane operating at the ENZ (upper) and FP (lower) resonant wavelengths.

To realize the CPA effect, next, we illuminate the nanowaveguides with two counter-propagating plane waves of equal intensity $I_0$ ( $I_0 = |\mathbf{E}_{in1}|^2 = |\mathbf{E}_{in2}|^2$ ) and different phases ($\psi_0$, $\psi_1$), as illustrated in Fig. 3(a). The transfer matrix method is used to investigate the role of interference from the two incident waves, which provides a relationship between the outgoing waves ($\mathbf{E}_{out1}$, $\mathbf{E}_{out2}$) and the input waves ($\mathbf{E}_{in1}$, $\mathbf{E}_{in2}$):

$$\begin{bmatrix} E_{out2} \\ E_{in2} \end{bmatrix} = M \begin{bmatrix} E_{in1} \\ E_{out1} \end{bmatrix} \quad (1)$$

where $M$ is the transfer matrix that connects the outgoing ($\mathbf{E}_{out1}$, $\mathbf{E}_{out2}$) and input ($\mathbf{E}_{in1}$, $\mathbf{E}_{in2}$) waves. In a reciprocal system, similar to our current case, the elements of the transfer matrix are related to the elements of the scattering matrix through the following relations [3]:

$$M = \begin{bmatrix} M_{11} & M_{12} \\ M_{21} & M_{22} \end{bmatrix} = \begin{bmatrix} t - \dfrac{r^2}{t} & \dfrac{r}{t} \\ -\dfrac{r}{t} & \dfrac{1}{t} \end{bmatrix} \quad (2)$$

where $t$ and $r$ represent the transmission and reflection coefficients under a single normal incident plane wave. CPA occurs when the outgoing waves from each side disappear, i.e., $E_{out1}=E_{out2}=0$, which means that the transfer matrix element $M_{11}$ becomes equal to zero ($M_{11}=0$), a relationship that can be directly derived by Eq. (1). Combining the relationship of $M_{11}$ with the scattering matrix elements given in Eq. (2), it implies that $t^2 = r^2$. Hence, in order to achieve CPA through destructive interference, $t$ and $r$ not only need to have equal amplitudes ($|t|=|r|$), but they must also have a correct phase difference $\Delta\phi$, being equal to either 0 or $\pm\pi$ [3]. Both these conditions can be perfectly satisfied at the nanoscale by using the proposed ENZ plasmonic waveguides system. Specifically, transmission is equal to reflection at $\lambda = 1204$ nm (as can be seen in Fig. 2), which is very close to the ENZ cut-off wavelength, and the phase difference between the two coefficients reaches to $-\pi$ at this frequency point. Therefore, CPA is obtained through destructive interference with two counter-propagating waves that have equal intensities and zero phase difference ($\Delta\psi = \psi_1 - \psi_0 = 0$) at this wavelength point (called the ENZ-CPA point). It is interesting that this effect can be achieved inside a lossless medium (glass loaded inside the nanochannels).

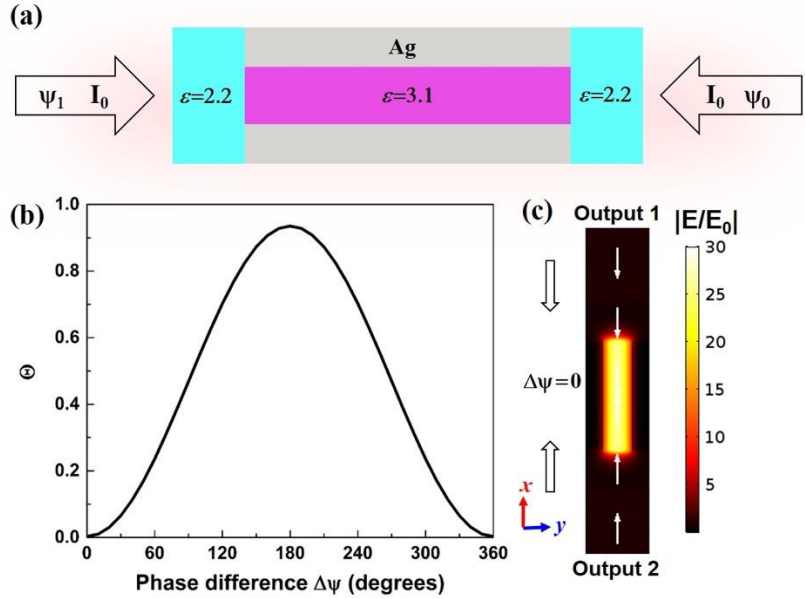

Figure 3. (a) Plasmonic waveguide illuminated by two plane waves from opposite directions. The two incident beams have equal intensity $I_0$ and different phases ($\psi_0$, $\psi_1$). (b) The computed output coefficient versus the phase difference of the two incident beams ($\Delta\psi = \psi_1 - \psi_0$). The wavelength is fixed at $\lambda = 1204$ nm, which lies in the vicinity of the ENZ resonance. (c) Normalized electric field distribution at the CPA point ($\Delta\psi = 0$). The white arrows indicate the total power flow directions. Strong and homogeneous electric field enhancement is obtained inside the ENZ plasmonic waveguides.

In order to quantitatively measure the amount of CPA, we define the following output coefficient $\Theta$: [10,12]

$$\Theta = \frac{|\mathbf{E}_{out1}|^2 + |\mathbf{E}_{out2}|^2}{|\mathbf{E}_{in1}|^2 + |\mathbf{E}_{in2}|^2} \quad (3)$$

that represents the ratio of total output power to the total input power impinging at the device. Therefore, $\Theta=0$ represents the CPA effect and $\Theta=1$ represents coherent perfect transmission. We calculate in Fig. 3(b) the output coefficient $\Theta$ as a function of the phase difference ($\Delta\psi = \psi_1 - \psi_0$) between the two incident waves. The wavelengths of the two incident waves are fixed at $\lambda = 1204$ nm, very close to the ENZ cut-off wavelength. Note that perfect CPA is achieved by using the phase offset $\Delta\psi = 0°$ or 360°, as we discussed before, while at $\Delta\psi = 180°$ almost total transmission can be obtained. Hence, the total output power can be modulated from perfect CPA to high transmission just by changing the phase difference $\Delta\psi$ between the two incident waves. Figure 3(c) demonstrates the computed normalized electric field distribution along one nanochannel when the phase offset is $\Delta\psi = 0°$ (CPA point). In this case, the entire incident radiation energy is totally absorbed by the waveguide channel and there is no reflection. The white arrows refer to the total power flow directions and clearly indicate the presented CPA effect close to the ENZ frequency. In addition, we note that the field enhancement almost doubles at the CPA point when illuminated by two waves compared with the case of just one incident wave shown in the right upper inset of Fig. 2. However, the field distribution is still uniform because the CPA wavelength is very close to the ENZ cut-off wavelength of the nanowaveguides. The obtained large and uniform field distribution at the ENZ-CPA point is advantageous in order to enhance coherent light-matter interactions at the nanoscale. For instance, it will result in large and uniform local density of optical states inside the nanochannels, an ideal condition to increase the collective spontaneous emission rate of several emitters placed inside the nanoslits leading to strong superradiance [23] and efficient entanglement [29].

## 3. NONLINEAR CPA WITH ENZ PLASMONIC WAVEGUIDES

The strong and uniform field enhancement inside the ENZ plasmonic waveguides can cause the proposed plasmonic system to access the optical nonlinear regime, especially for high incident wave intensities. In this section, we demonstrate that the third-order (Kerr) nonlinear effect will make the CPA effect intensity-dependent and tunable with ultrafast (femtosecond scale) speed.

Again, we illuminate the nonlinear ENZ plasmonic waveguides with two counter-propagating plane waves, as shown in Fig. 3(a). Different from the linear case, here we explore the third-order nonlinear permittivity of the dielectric slits, where the relative permittivity of glass $\varepsilon_1 = 3.1$ in section 2 (Fig. 1) is now changed to its nonlinear counterpart $\varepsilon_1' = 3.1 + \chi^{(3)} |\mathbf{E}_{ch}|^2$. The coefficient $\chi^{(3)} = 4.4 \times 10^{-20} \, \mathrm{m^2/V^2}$ consists typical third-order nonlinear susceptibility values of glass [30] and $\mathbf{E}_{ch}$ is the local electric field induced inside the nanoslits. We vary the phase difference $\Delta\psi$ between the two incident waves from 0° to 360° for a given incident wavelength at $\lambda = 1204$ nm (*i.e.,* the linear ENZ-CPA point) and record the maximum and minimum values of the output coefficient $\Theta$ as a function of the input intensity of each beam. The result is shown in Fig. 4(a). Note that as we gradually increase the input intensity before the threshold value of 100MW/cm$^2$, the minimum $\Theta$ values always stay very low and near zero, meaning that the CPA effect prevails under the illumination of small input intensities. The minimum $\Theta$ values rapidly increase above the relative small input intensity threshold value and gradually merge with the maximum $\Theta$ values. Eventually, the minimum $\Theta$ values reach to almost 1, which indicate that the CPA effect gradually disappears due to the nonlinear response of the structure. In addition, the output power eventually becomes phase-insensitive just by increasing the input intensity. Figure 4(b) illustrates the change in the output coefficient $\Theta$ versus the phase difference $\Delta\psi$ at very low (black line) and high (red line) input intensities. The Kerr nonlinear effect is relatively weak for the case of low input intensity $I_0 = 0.1 \, \mathrm{MW/cm^2}$, resulting that the effective permittivity $\varepsilon_1'$ is still very close to its linear value $\varepsilon_1 = 3.1$. Therefore, the computed output coefficient (black line) is very similar to the one obtained for the linear case [Fig. 3(b)] and CPA can be achieved by using phase offsets of $\Delta\psi = 0°$ or 360°, as it was discussed in the previous section. However, the computed output coefficient values (red line) become large and phase insensitive in the case of higher input intensity $I_0 = 3000\,\mathrm{MW/cm^2}$, indicating that the third-order nonlinearity has strongly influenced the effective permittivity of glass and the CPA conditions are not satisfied anymore. This tunable intensity-dependent CPA effect is similar to the mechanism of nonlinear saturable absorption [30]. Such strong nonlinear tunable CPA response, achieved with relative low input intensities and at the nanoscale, is a unique feature of the proposed nonlinear ENZ plasmonic structure. Note that photonic CPA configurations, which are the usually proposed in the literature, cannot achieve such strong nonlinear response due to poor field concentration and enhancement inside their geometries.

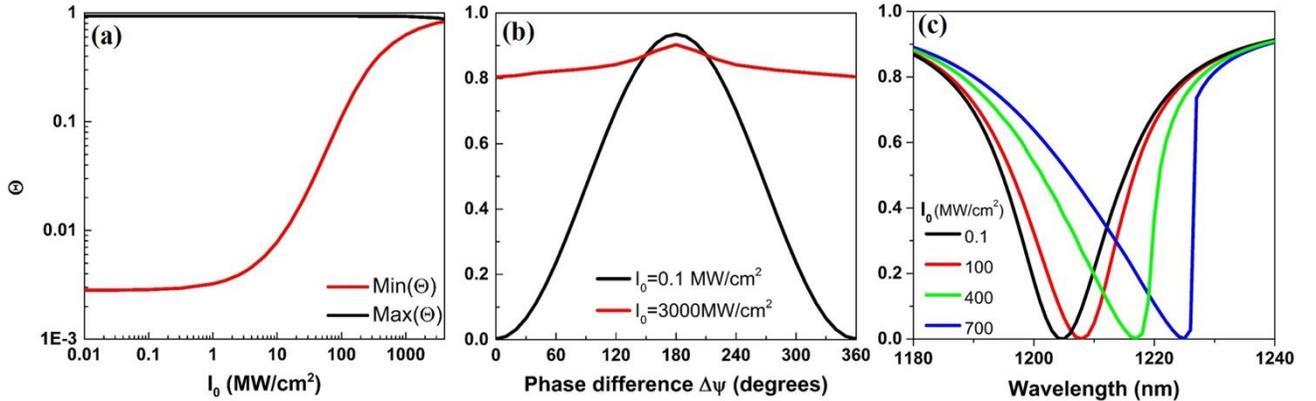

Figure 4. (a) Nonlinear line shapes of the maximum and minimum values of the output coefficient Θ versus the input intensity $I_0$ of each beam. (b) Output coefficient Θ versus the phase difference for two input intensity values: $I_0$=0.1 MW/cm$^2$ and $I_0$=1000 MW/cm$^2$. The wavelength is fixed to 1204 nm (close to ENZ) for both (a) and (b) cases. (c) Output coefficient Θ versus the incident wavelength in the vicinity of the ENZ resonance for four different input intensity values: $I_0$= 0.1 MW/cm$^2$, 100 MW/cm$^2$, 400 MW/cm$^2$ and 700 MW/cm$^2$.

Next, we plot the output coefficient Θ versus the incident wavelength around the ENZ resonance for four different input intensities and the results can be seen in Fig. 4(c). In this scenario, the phase difference $\Delta\psi$ is fixed at 0° meaning that the computed Θ values in this figure reflect the minimum output coefficient or the CPA condition. The wavelength that CPA occurs is $\lambda = 1204$ nm for a low input intensity $I_0 = 0.1\,\mathrm{MW/cm^2}$, same as the linear ENZ-CPA wavelength. Interestingly, the perfect CPA effect always exists, as we increase the input intensity $I_0$ from $0.1\,\mathrm{MW/cm^2}$ to $700\,\mathrm{MW/cm^2}$, but the absorption dip experiences a redshift towards longer wavelengths. It should also be noted that for $I_0 = 700\,\mathrm{MW/cm^2}$ an abrupt transition of Θ is obtained, indicating that the presented ENZ system will exhibit optical bistable response for this and higher input intensity values [20].

The presented nonlinear processes usually require high laser powers in order to be excited, which can cause damage to the proposed ENZ plasmonic nanostructures. Hence, it is interesting to check the influence of the thermal effects to the metallic parts of the proposed nonlinear ENZ waveguides when the input intensity values are increased. This will provide further insights in terms of the practicality of the presented tunable CPA scheme. To this end, we introduce the heat transfer equation in our nonlinear electromagnetic simulations, which are performed using the commercial finite-element simulation software COMSOL Multiphysics. The related thermal parameters used in our simulations follow previously derived experimental data [31]. Based on the revised multiphysics (electromagnetics plus thermal) simulation model, we compute and plot in Figs. 5(a)-(b) the maximum and average temperature values induced at the plasmonic waveguide around the ENZ wavelength for two different input intensities. In the relative low input intensity case ($I_0$=100 MW/cm$^2$), there is no substantial temperature increase in the metallic parts and both maximum and average temperatures are almost equal to the surrounding room temperature (293 K). However, the temperature increases when the ENZ waveguide is inset: Computed temperature distribution at the ENZ CPA wavelength point (1228 nm) for high input intensity $I_0$=1000 MW/cm$^2$.

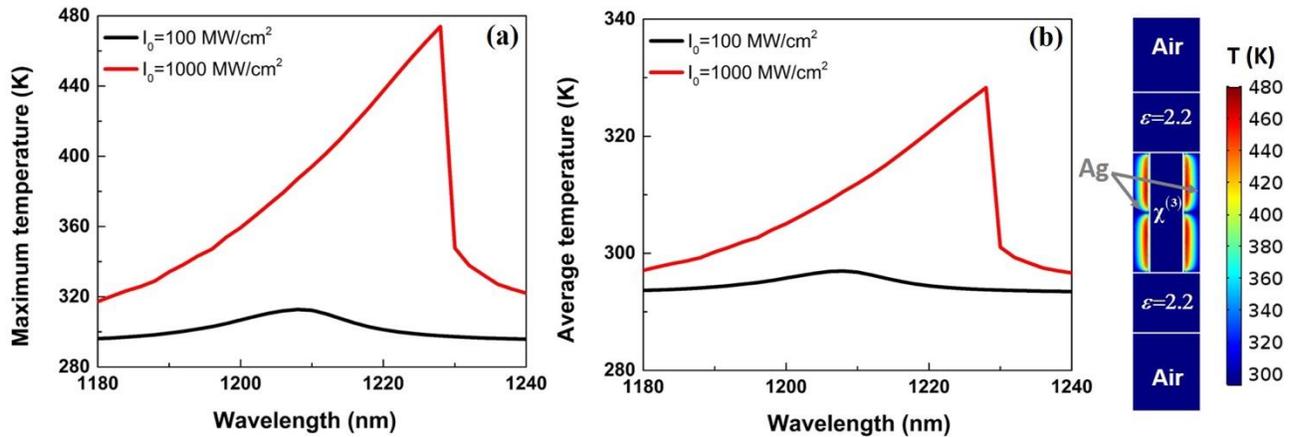

Figure 5. Computed maximum (a) and average (b) induced temperature values versus the incident wavelength around the ENZ resonance for two input intensities with low $I_0$=100 MW/cm$^2$ and high $I_0$=1000 MW/cm$^2$ input intensity values. Right

illuminated by high input intensities ($I_0$=1000 MW/cm$^2$) counter-propagating waves, as it can be clearly seen in Fig. 5. The corresponding temperature distribution at the ENZ CPA point (1228 nm) is shown in Fig. 5(c) when two counter-propagating waves with equal and high input intensities $I_0$=1000 MW/cm$^2$ illuminate the structure. It can be clearly seen that the temperature is increased only along the lossy silver screen parts and not in the lossless dielectric slit, superstrate, and substrate, as it was expected. Interestingly, the maximum temperature values are only increased by approximately 180 K. These values are still much lower than the silver's melting point (1235 K). Therefore, the relative small increase in the temperature is not expected to trigger any unwanted slow thermal nonlinearities or cause damage to the metallic materials (melting). Hence, the proposed nonlinear tunable ENZ structures will only be characterized by the ultrafast (fsec-scale) electronic nonlinearities.

## 4. CONCLUSIONS

We proposed an efficient way to enhance coherent and nonlinear light-matter interactions at the nanoscale based on ENZ plasmonic waveguides. The proposed plasmonic screens are practical and can be fabricated by embedding them in a dielectric material superstrate and substrate. This device behaves as an effective ENZ material at the cut-off resonance wavelength resulting in an anomalous impedance matching phenomenon that is able to produce large and uniform field enhancement along its nanochannels. When we illuminate the waveguide with two counter-propagating waves, the CPA effect was found to exist around the ENZ cut-off wavelength leading to double field enhancement inside the nanochannels in comparison to the single incident wave case. The strong and uniform field enhancement at the ENZ-CPA point can cause the plasmonic system to access the optical nonlinear regime exhibiting boosted third-order optical nonlinear effects at the nanoscale. As a result, the phase-dependent CPA effect also becomes intensity-dependent and tunable with ultrafast speed due to the third-order nonlinear susceptibility of the dielectric nanochannels. In addition, the influence of the thermal effects in the proposed nonlinear ENZ waveguides is investigated, as we increase the input incident waves' intensities. It is found that the induced temperature increase is minor and is not expected to affect the performance of the proposed ENZ CPA devices. This is mainly due to the relative low input intensities used to excite the proposed structures and trigger nonlinear effects. We believe that our results can provide a new platform to design optical switches [15], nonlinear gap solitons [14], and tunable unidirectional coherent perfect absorbers [27].

## ACKNOWLEDGMENTS

The work has been partially supported by National Science Foundation (NSF) (DMR-1709612) and Nebraska Materials Research Science and Engineering Center (MRSEC) (DMR-1420645).